\newcommand{\bluecov}{{BlueCov}}
\newcommand{\jbmc}{JBMC}
\newcommand{\diffblue}{{Diffblue Ltd}}

\newcommand{\artifacturl}{\url{https://www.dropbox.com/s/a7pe39ygj5znuh4/bluecov.zip}}

\def\BibTeX{{\rm B\kern-.05em{\sc i\kern-.025em b}\kern-.08em
    T\kern-.1667em\lower.7ex\hbox{E}\kern-.125emX}}

\newcommand{\hide}[1]{}

\documentclass[sigconf,]{acmart}

\usepackage[utf8]{inputenc}  
\usepackage[T1]{fontenc}     
\usepackage{microtype}       
\usepackage[english]{babel}  
\usepackage{graphicx}        
\usepackage{comment}         
\usepackage{flushend}        
\usepackage{listings}

\setcopyright{none}

\setcopyright{none}
\renewcommand\footnotetextcopyrightpermission[1]{} 
\usepackage{algorithm}
\usepackage{algorithmicx}
\usepackage{algpseudocode}

\usepackage{subfig}

\usepackage{xcolor}
\usepackage{todonotes}

\graphicspath{{fig/}}

\begin{document}

\title{BlueCov: Integrating Test Coverage and Model Checking with JBMC}
\author{Matthias Güdemann}
\affiliation{
  \institution{Munich University of Applied Sciences HM, Germany}
  \country{Germany}
}
\author{Peter Schrammel}
\affiliation{
  \institution{University of Sussex and Diffblue Ltd., UK}
  \country{United Kingdom}
}
\begin{abstract}
Automated test case generation tools help businesses to write tests and increase
the safety net provided by high regression test coverage when making code
changes. Test generation needs to cover as much as possible of the uncovered
code while avoiding generating redundant tests for code that is already covered
by an existing test-suite.

In this paper we present our work on a tool for the real world application of
integrating formal analysis with automatic test case generation. The test case
generation is based on coverage analysis using the Java bounded model checker
(JBMC). Counterexamples of the model checker can be translated into Java method
calls with specific parameters.

In order to avoid the generation of redundant tests, it is necessary to measure
the coverage in the \emph{exact same} way as JBMC generates its coverage
goals. Each existing coverage measurement tool uses a slightly different
instrumentation and thus a different coverage criterion. This makes integration
with a test case generator based on formal analysis difficult. Therefore, we
developed \bluecov{} as a specific runtime coverage measurement tool which uses
the exact same coverage criteria as JBMC does. This approach also allows for
incremental test-case generation, only generating test coverage for previously
untested code, e.g., to complete existing test suites.
\end{abstract}

%


%
%


\maketitle



\section{Introduction}
\label{sec:introduction}

In recent years there has been steady progress in developing automated
formal verification tools. Promising tools have emerged, not only
targeting the traditional domain of embedded C programs, but also for
languages such as Java that are predominantly used for larger
enterprise software systems~\cite{cordeiro2018jbmc}.
For such systems, testing is the primary way of obtaining confidence
in the correct functioning of the system~\cite{DBLP:conf/fmcad/Schrammel20a}.
While it is far from feasible to formally verify such systems,
automated verification tools can be used to verify small subsystems
or support the automation of the testing process.
An important use case is to automatically generate tests for untested parts of
the system~\cite{DBLP:conf/vmcai/HolzerSTV09,DBLP:conf/nfm/NellisKCKST16}.
For that purpose, information about uncovered code needs to be communicated
to the verification tool in order to reduce the computational effort and
avoid generation of redundant tests.

In this paper, we present the \bluecov{} tool which solves this
communication problem in practice for Java programs using the bounded model
checker, \jbmc~\cite{cordeiro2018jbmc}.
Similar ideas may apply to other programming languages.

In Java, the imperative code is first compiled into a stack-based
language, the \emph{Java bytecode}, which is interpreted by the Java
Virtual Machine (JVM).
This compilation step abstracts away a great deal of syntactic
sugar and language-level complications. Also, libraries which
make up the vast majority of any real-world Java application
are supplied and linked on the bytecode level.
Therefore, verification tools also perform their analysis based
on bytecode as input.
The stack-based bytecode is translated into a
control flow graph (CFG) to perform further analysis, such as
abstract interpretation or symbolic execution.

Code coverage measurement is performed by bytecode-level instrumentations. This
information is used to avoid generating tests for code that is already covered
by an existing test-suite. The problem is now that the code structure on the
bytecode level differs from the CFG used by the verification tool in details
that are relevant for communicating structural code coverage information.
Hence, it is not straightforward to use test coverage information
obtained from coverage measurement tools, such as JaCoCo~\cite{jacoco},
in a verification tool.
%
An exact correspondence is required, however, in order to satisfy the
desired property of the analysis such as pruning paths and avoiding
the generation of redundant tests.

One way of achieving this is to pick a coverage measurement tool
and implement the exact same coverage criterion inside the test generator,
including all the oddities of the coverage measurement tool and
keep it up-to-date as the tool changes.
If we wanted to support further coverage criteria, e.g.\ modified condition /
decision coverage (MC/DC), then we would need to find an appropriate mapping
from the existing coverage instrumentation or even modify its instrumentation if
it does not capture all information required.

We took a more maintainable and flexible approach here by developing \bluecov{},
which takes the coverage goals \emph{from the test generator} and performs the
instrumentation accordingly. The coverage criterion is hence defined by the test
generator, i.e., \jbmc{} in our case.

\bluecov{} is much simpler than existing coverage measurement tools, also
because it does not require computing coverage percentages.  Coverage
percentages are highly dependent on the exact coverage criterion and developers
are often confused by mismatches between numbers reported from different
tools. We therefore do not use \bluecov{} for reporting coverage percentages,
but only for communication with the test generator. Originally, \bluecov{} was
developed for and successfully used in the \diffblue{} Cover commercial test case
generator. It has now been
open-sourced\footnote{\url{https://github.com/diffblue/BlueCov}}.

This paper is an extended version of~\cite{bluecovFullVersion}. It is structured
as follows: Section~\ref{sec:jbmc-java-bounded} gives some background on \jbmc{}
and coverage analysis, Section~\ref{sec:bluecov} explains the combination of
\jbmc{} and \bluecov{}, Section~\ref{sec:example} illustrates the approach with
an example, Section~\ref{sec:bytec-instr} gives details on the implementation
and Section~\ref{sec:conclusion} concludes the paper.

\section{Background}
\label{sec:jbmc-java-bounded}

\subsection{\jbmc --- Java Bounded Model Checking}
\label{sec:jbmc-java-bounded-1}

\jbmc~\cite{cordeiro2018jbmc,DBLP:conf/tacas/CordeiroKS19} is a bounded model
checker for Java which uses the compiled class files as input.  In this sense it
could also analyze other languages that are compiled to Java bytecode, e.g.,
Kotlin. It can also read jar files which are a form of structured zip archive of
class files.


\jbmc{} is a front-end to the CProver framework. CProver is also used by
the CBMC model checker for C~\cite{kroening2014cbmc}. \jbmc{} translates
the Java bytecode into CProver's internal GOTO program code. GOTO is a
simple sequential programming language which is very close to C in its
execution model.  It represents the CFG of the program using GOTO
instructions to model CFG edges.  Java bytecode on the other hand is a
stack-based low-level, assembly code like programming language.

Only necessary classes and functions are translated to GOTO.\@ For this, \jbmc{}
performs an over-approximate reachability analysis on the call graph. Only
functions which are actually used are added to the GOTO program.




\jbmc{} uses the SMT solver built into CProver as backend, which implements a
reduction of bitvector, floating-point, array and string theories to SAT.
\jbmc{} competes in the Java track at the Software Verification Competition
(SV-COMP)~\cite{DBLP:journals/sigsoft/CordeiroKS18,DBLP:conf/tacas/Beyer19}, has
won once and was among the top three tools otherwise.

\subsection{Coverage Analysis in \jbmc{}}
\label{sec:cover-analys-jbmc}

For test case generation \jbmc{} currently uses the \emph{location} coverage
criterion. This criterion tries to cover each bytecode in the class file.  For
each CFG node corresponding to distinct bytecode instructions, \jbmc{} inserts an
assertion node \textsf{assert(false)} representing the coverage goal into the
CFG.\@ An analysis with \jbmc{} then tries to deduce values for the input
parameters of the method in such a way that the assertions are reached. These
synthesized input values are then translated into executable test cases. For
\bluecov{} other coverage criteria are supported, too, e.g. branch, path or
MC/DC coverage.

The translation from \jbmc{} results to executable Java is not
trivial. Primitive values are relatively easy, but complex objects can
be difficult to create. This is true in particular when one does not
want to use reflection to create object instances. While reflection is
very powerful, it generally does not correspond to readable tests a
human tester would write. Reflection also allows for creation of
object states which could not be created programmatically, e.g.,
because internal invariants are not respected. The Java code test case
generation is proprietary to the company \diffblue{} Ltd.\ and thus not
part of the presentation here.

\subsection{Testcase Minimization}
\label{sec:testc-minim}

In many applications \jbmc{} will have to create multiple new test cases in order
to complete the test suite. Often there are different possibilities to cover the
remaining coverage goals. As each generated unit test might cover several
coverage goals, it makes sense to try to minimize the number of generated unit
tests to complete the coverage.

The underlying problem is the subset cover problem which is
$\mathcal{NP}$-complete and can therefore be expensive to solve if the
number of tests is large. \jbmc{} uses a greedy heuristic to approximate
the minimal number of new unit tests. First, traces are sorted by the
number of covered goals. Traces are then processed in this order,
collecting newly covered goals and dropping those traces that do not
cover any new goal. This is similar to the approach used
by~\cite{DBLP:journals/sttt/SchrammelMK16}.

This approximation works well in
practice to reduce the number of generated test cases.


\subsection{Java Specific Challenges}
\label{sec:java-spec-chall}

Java is an object-oriented programming language with an extensive standard
library. This poses some additional challenges for model-checking of Java
programs as compared to C programs. We give some challenges here and illustrate
how \jbmc{} handles these. These challenges can have a big impact on differences
between reported and measured coverage because \jbmc{} might have a different model
than what is really executed.

\textbf{Object Orientation} Java methods often use the interfaces as input
parameters. As interfaces do not have an implementation, an analysis tool has to
use a class which implements the interface. This often requires loading more
classes than the ones that are actually referenced in a class file. For example,
using the \emph{List} interface would create a reference to \emph{List} but
would also require loading another class that implements the interface, e.g.,
\emph{ArrayList}.

Another challenge is the use of generics in the class file. Java uses type
erasure which means that there is no typing information in the class files. In
such case \jbmc{} analyzes the bytecode instructions for explicit casts without
checks. As the compiler does have the necessary information, it can emit
instructions that do not check the result after a
cast.  

\textbf{Java Class Library} An important challenge when analyzing Java is the
Java class library (JCL). It provides different data structures and
algorithms. It is not feasible to simply load the JCL class files, as the
resulting model would be much too big. What \jbmc{} does is to provide a \emph{jar}
file that contains \emph{models} of commonly used parts of the JCL.\@ Such
models do not implement the respective class functionality directly but contain
a sufficient specification for \jbmc{}.

In the running example (cf. listing~\ref{lst:example-program}) in this paper,
the \emph{core models} library is used to find the model for the \emph{Math.abs}
function. In this case the implementation is the same as the original
implementation in the JCL as shown in listing~\ref{lst:math-abs}.

\begin{lstlisting}[caption=\texttt{Math.abs} Implementation\label{lst:math-abs}]
public static float abs(float a) {
    return (a <= 0.0F) ? 0.0F - a : a;
}
\end{lstlisting}

For other functions \jbmc{} provides a different implementation which is
functionally equivalent to the original one, but is easier for the SMT solver to
handle. An example for this is the \emph{max} (maximum) function for
\emph{float} parameters. The \jbmc{} core models version of this function is
shown in listing~\ref{lst:math-max}. The commented-out section contains the
original code of the JCL.\@ The \jbmc{} implementation is functionally
equivalent according to the \textsc{Ieee}754 standard but is simpler for \jbmc{}
to analyze. For example, it does not use the \emph{Float.floatToRawIntBits}
method.

\begin{lstlisting}[caption=Core Model version of \texttt{Math.max}\label{lst:math-max}]
public static float max(float a, float b) {
  // original code
  // if (a != a)
  //     return a;   // a is NaN
  // if ((a == 0.0f) &&
  //     (b == 0.0f) &&
  //     (Float.floatToRawIntBits(a)
  //          == negativeZeroFloatBits)) {
  //     // Raw conversion ok since NaN
  //     // can't map to -0.0.
  //     return b;
  // }
  // return (a >= b) ? a : b;

  // JBMC core-models implementation
  if (Float.isNaN(a) || Float.isNaN(b)) {
      return Float.NaN;
  } else {
    float result = CProver.nondetFloat();
    // choose result in such a way that it is
    // the maximum of a and b

    CProver.assume( (result == a || result == b)
                 && result >= a && result >= b);
    return result;
  }
}
\end{lstlisting}

Some JCL functions are currently not implemented or modeled. These functions
simply return a nondeterministic value of the correct type when called. If this
is a problem for an analysis, one can additionally model the required function
which increases the precision of the analysis and of the test creation done by
\jbmc{}.

\section{Combining \jbmc{} and \bluecov{} }
\label{sec:bluecov}

As explained above, there exist several tools which report coverage of Java
programs, but none uses precisely the same criteria as \jbmc{} does. For
real-world project, it is an important goal to minimize the effort to complete
the test suite. This requires measuring the test coverage in the exact way as
\jbmc{} does. Therefore, the \bluecov{} tool was developed to facilitate
integration with \jbmc{}.\@ It tracks the exact Java bytecode instructions that
it associates with its coverage criterion. \jbmc{} then provides this
information to \bluecov{}, which instruments the same bytecode instructions for
runtime coverage measurement. That way, we can close the gap between the
\emph{execution coverage} that can be obtained from test execution and the
\emph{generation coverage} that \jbmc{} reports to have achieved during test
generation.

The coverage goals determined by \jbmc{} to drive its test generation are
reported as JSON output. For each coverage goal \jbmc{} emits a
Boolean flag covered or not covered. This is called the
\emph{generation coverage} of \jbmc{}.\@ \jbmc{} also calculates values for
the input parameters for Java methods in order to reach the coverage
goals.

\begin{figure*}[ht]
  \centering
  \includegraphics[width=0.5\textwidth]{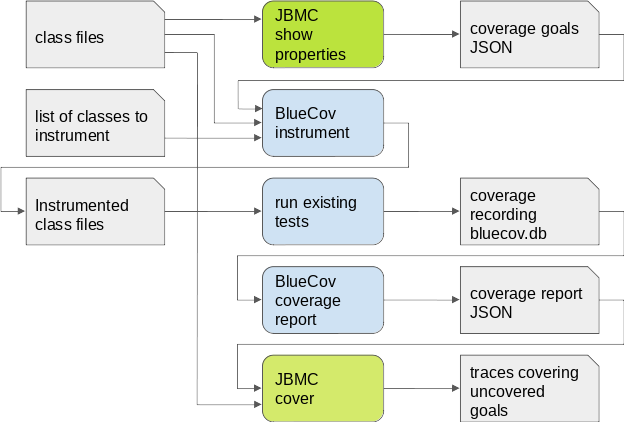}
  \caption{Overview of the Approach (\jbmc{} --- green, \bluecov{} --- blue)}
  \label{fig:bluecov-overview}
\end{figure*}

If there is a mismatch in the understanding of the coverage criteria between the
test generator and the coverage measurement tool then the generation coverage
might be different from the \emph{execution coverage}. The execution coverage is
the coverage which is reached when the code is actually executed on the JVM
using the input parameters calculated by \jbmc{}.\@
Figure~\ref{fig:bluecov-overview} shows the overall approach.

\textbf{Class Files to Properties} The first step is to give the class files of
the project under analysis to \jbmc{}.\@ It provides the option to show the
coverage goals (or \emph{properties}) considered for \emph{generation
  coverage}. The output is given in JSON format and contains all the information
necessary for \bluecov{} to perform the bytecode instrumentation.

\textbf{Property Instrumentation} In the second step the \bluecov{} tool takes
the output of the \jbmc{} coverage goals, the class files of the project and a list
of class files to instrument as its input. It then proceeds by creating a
database file and instruments the class files with each coverage goal.

\textbf{Measure Existing Coverage} In the third step the existing test suite of
the project is run using the instrumented class files. During this run,
\bluecov{} registers when an instrumented coverage goal is reached and increases
a counter in the database file for each reached coverage goal.

\textbf{Coverage Report} After the run of the test suite, \bluecov{} reports the
measured \emph{execution coverage}. The report is in JSON format and contains
the hitcount for each coverage goal.

\textbf{Enhance Test Coverage} The \bluecov{} coverage report is then used as
input for \jbmc{} create a minimal number of additional tests to complete the
coverage of the existing test suite. \jbmc{} reports the \emph{generation coverage}
that it achieved together with the generated test inputs to the methods.

\section{Example using \bluecov{}}
\label{sec:example}

We use the example in listing~\ref{lst:example-program} to illustrate the steps
of our approach based on \bluecov. The below function \emph{FloatTools.sign}
returns an \emph{int} depending on the input \emph{float} parameter. The intent
is to have a sign function of the input parameter \emph{x}. The first case in
line 3 covers the ``zero'' value, i.e., a very small absolute value of \emph{x}
and returns $0$. The second case (line~5) returns $-1$ in case \emph{x} is
negative and the third case (line~7) returns $1$ in case \emph{x} is
positive. The return statement in line~9 was added because the compiler would
otherwise emit the error \emph{error: missing return statement}.

\begin{lstlisting}[caption=Example Program\label{lst:example-program}]
public class FloatTools {
  public static int sign(float x) {
    if(Math.abs(x) < 1e-6)
      return 0;
    if(x < 0)
      return -1;
    if (x > 0)
      return 1;
    return -2;
  }
}
\end{lstlisting}

The existing test suite is shown in listing~\ref{lst:test-example-program},
i.e., testing each of the three cases:

\begin{lstlisting}[caption=Test Suite for Example Program\label{lst:test-example-program}]
public class FloatToolsTest {
  @Test
  public void testSignZero () {
    assertEquals(0, FloatTools.sign(-1e-10f));
  }
  @Test
  public void testSignNeg () {
    assertEquals(-1, FloatTools.sign(-10f));
  }
  @Test
  public void testSignPos () {
    assertEquals(1, FloatTools.sign(1234f));
  }
}
\end{lstlisting}

\subsection{Class files to Properties}
\label{sec:class-to-properties}

\jbmc{} converts the bytecode instructions of \emph{FloatTools.sign} into
its internal GOTO representation and inserts coverage goals according
to the given coverage criterion. For \emph{location} coverage, \jbmc{}
creates 9 location coverage goals for \emph{FloatTools.sign}
corresponding to the \texttt{ASSERT} statements in the GOTO function
below.%
\footnote{
  Simplified to improve readability;
  the full output can be produced with the artifact available at
  \artifacturl{}
}

\begin{lstlisting}[basicstyle=\scriptsize,xleftmargin=-0.5ex,numbers=none]
FloatTools.sign(float) /* FloatTools.sign:(F)I */
  DECL return_tmp0 : floatbv[32]
  // FloatTools.java line 5 FloatTools.sign:(F)I bytecode-index 1
  ASSERT false // goal FloatTools.sign:(F)I.coverage.1
  // FloatTools.java line 5 FloatTools.sign:(F)I bytecode-index 1
  CALL java.lang.Math.<clinit>()
  // FloatTools.java line 5 FloatTools.sign:(F)I bytecode-index 1
  CALL java.lang.Math.abs:(F)F(x)
  // FloatTools.java line 5 FloatTools.sign:(F)I bytecode-index 1
  ASSIGN return_tmp0 := java.lang.Math.abs:(F)F::return_value
  // FloatTools.java line 5 FloatTools.sign:(F)I bytecode-index 1
  DEAD java.lang.Math.abs:(F)F::return_value
  // FloatTools.java line 5 FloatTools.sign:(F)I bytecode-index 5
  ASSERT false // goal FloatTools.sign:(F)I.coverage.2
  // FloatTools.java line 5 FloatTools.sign:(F)I bytecode-index 5
  IF (NOT isnan(return_tmp0)) AND
      (return_tmp0 < 1.000000e-6)) THEN GOTO 1
  // FloatTools.java line 7 FloatTools.sign:(F)I bytecode-index 8
  ASSERT false // goal FloatTools.sign:(F)I.coverage.3
  // FloatTools.java line 7 FloatTools.sign:(F)I bytecode-index 8
  DEAD return_tmp0
  // FloatTools.java line 7 FloatTools.sign:(F)I bytecode-index 8
  GOTO 2
  1: DEAD return_tmp0
  // FloatTools.java line 6 FloatTools.sign:(F)I bytecode-index 7
  ASSERT false // goal FloatTools.sign:(F)I.coverage.4
  // FloatTools.java line 6 FloatTools.sign:(F)I bytecode-index 7
  ASSIGN FloatTools.sign:(F)I::return_value := 0
  // FloatTools.java line 6 FloatTools.sign:(F)I bytecode-index 7
  GOTO 5
  // FloatTools.java line 7 FloatTools.sign:(F)I bytecode-index 11
  2: ASSERT false // goal FloatTools.sign:(F)I.coverage.5
  // FloatTools.java line 7 FloatTools.sign:(F)I bytecode-index 11
  IF isnan(x) OR (x >= 0) THEN GOTO 3
  // FloatTools.java line 8 FloatTools.sign:(F)I bytecode-index 13
  ASSERT false // goal FloatTools.sign:(F)I.coverage.6
  // FloatTools.java line 8 FloatTools.sign:(F)I bytecode-index 13
  ASSIGN FloatTools.sign:(F)I::return_value := -1
  // FloatTools.java line 8 FloatTools.sign:(F)I bytecode-index 13
  GOTO 5
  // FloatTools.java line 9 FloatTools.sign:(F)I bytecode-index 17
  3: ASSERT false // goal FloatTools.sign:(F)I.coverage.7
  // FloatTools.java line 9 FloatTools.sign:(F)I bytecode-index 17
  IF isnan(x) OR (x <= 0) THEN GOTO 4
  // FloatTools.java line 10 FloatTools.sign:(F)I bytecode-index 19
  ASSERT false // goal FloatTools.sign:(F)I.coverage.8
  // FloatTools.java line 10 FloatTools.sign:(F)I bytecode-index 19
  ASSIGN FloatTools.sign:(F)I::return_value := 1
  // FloatTools.java line 10 FloatTools.sign:(F)I bytecode-index 19
  GOTO 5
  // FloatTools.java line 11 FloatTools.sign:(F)I bytecode-index 21
  4: ASSERT false // goal FloatTools.sign:(F)I.coverage.9
  // FloatTools.java line 11 FloatTools.sign:(F)I bytecode-index 21
  ASSIGN FloatTools.sign:(F)I::return_value := -2
  5: END_FUNCTION
\end{lstlisting}

As Java bytecode is used for the analysis, there can be multiple
coverage goals per source code line. The coverage goals map to the
original source code as shown in listing~\ref{lst:goal-mapping}.
\footnote{
  For readability, we write \texttt{goal 1} for the goal with name \texttt{FloatTools.sign:(F)I. coverage.1}.
}

\begin{lstlisting}[caption=Mapping of Goals to Source Code\label{lst:goal-mapping}]
public class FloatTools {
  public static int sign(float x) {
    if(Math.abs(x) < 1e-9) // goal 1, goal 2
      return 0;            // goal 4
    if(x < 0)              // goal 3, goal 5
      return -1;           // goal 6
    if (x > 0)             // goal 7
      return 1;            // goal 8
    return -2;             // goal 9
  }
}
\end{lstlisting}

\jbmc{} outputs information about the coverage goals in JSON format. For example,
for goal \texttt{FloatTools.sign:(F)I.coverage.1} it produces an entry as shown
below. This also includes information to map the bytecode information to the
line information of the source Java program.

\begin{lstlisting}
{
  "class": "coverage",
  "coveredLines": "5",
  "description": "block 2 (lines FloatTools.java:5)",
  "expression": "false",
  "name": "FloatTools.sign:(F)I.coverage.1",
  "sourceLocation": {
    "bytecodeIndex": "1",
    "file": "FloatTools.java",
    "function": "FloatTools.sign:(F)I",
    "line": "5"
  }
}
\end{lstlisting}

Analogous entries exist for the other eight coverage goals. Each has a unique
name and a source location which it covers. For Java, this source location
contains the bytecode index of the goal, the class name, the method name and the
parameter and return value types.
\footnote{ The full output can be produced using the artifact available at
  \artifacturl}

\subsection{Property Instrumentation}
\label{sec:property-instrumentation}

Using the information about the goals as generated by \jbmc{}, \bluecov{}
instruments the class files. It either creates an empty coverage database file
or adds new entries to an existing one.
The default location and name is \emph{blueCov.db}. For each of the generated
goals there is a unique identifier used to identify the goal in the database
file. This UID is constructed from the fully qualified name, function parameter
types and bytecode index, e.g., \texttt{FloatTools.sign:(F)I@1} for UID 0.

The bytecode instructions of the \emph{sign}
method first call the \emph{Math.abs} function which takes a
\emph{float} parameter and returns a value of type \emph{float}.  This
is shown below, first the original code then the instrumented code.

\begin{lstlisting}
public static int sign(float); // original code
  Code:
     0: fload_0
     1: invokestatic  #2
     // Method java/lang/Math.abs:(F)F
     ...

public static int sign(float); // instrumented code
  Code:
    0: fload_0
    1: getstatic     #17
    // Field company_coverage_reporter:
    //            Lorg/cprover/coverage/CoverageLog;
    4: ldc           #18 // int 0
    6: invokevirtual #24
    // Method org/cprover/coverage/CoverageLog.record:(I)V
    9: invokestatic  #30
    // Method java/lang/Math.abs:(F)F
\end{lstlisting}

In the instrumented code first the static instance of \emph{CoverageLog} is
loaded, then its \emph{record} method is called with the UID $0$. This
identifier corresponds to the above goal. This call increments the hitcount for
this goal in the \bluecov{} in-memory database.

The class itself does not contain the \emph{clinit} static initializer as it
does not contain any static fields. A static field is needed by \bluecov{} and
therefore it adds a static initializer to get the singleton
\emph{org.cprover.coverage.CoverageLog} object used for logging. The generated
bytecode instructions of the static initializer looks as follows:

\begin{lstlisting}
static {};
  Code:
    0: invokestatic  #43
    // Method org/cprover/.../
    //          CoverageLog.getInstance:()Lorg/.../
    //          CoverageLog;
    3: putstatic     #15
    // Field company_coverage_reporter:
    //         Lorg/cprover/coverage/CoverageLog;
    6: return
\end{lstlisting}

Each of the other coverage goals is instrumented analogously. Each
instrumentation uses three additional bytecode instructions, each with a
different UID for the database. The UID is loaded as a parameter using the
\emph{ldc} or load constant bytecode instruction.  \footnotemark{}

This code instrumentation causes a constant time overhead when executing such
instrumented bytecode. This could be a problem in really long-running unit
tests, in particular within loops. If the exact number of hits is not required,
then a further optimization is possible which records only the first time a
specific location is covered and skips logging afterwards. This effectively
reduces the runtime overhead to a branch execution that can always be predicted
correctly after the first time.

\subsection{Measure Existing Coverage}
\label{sec:measure-coverage}

To measure the existing coverage, the \emph{record} method of the
\emph{CoverageLog} class is called each time a coverage goal is reached. When
called the first time for a specific coverage goal, the method inserts the key
with value 1 into the hash map. At each further call, the associated value is
incremented, updating the hitcount entries in the database.  For this,
\bluecov{} needs to be added to the Java classpath.%
\footnotemark[\value{footnote}]

\subsection{Coverage Report}
\label{sec:coverage-report}

After the test suite is finished, the coverage reporter of \bluecov{} is called
which reports the measured coverage. The reporter iterates over all entries in
the database and emits the measured hitcount for each coverage goal.%
\footnotemark[\value{footnote}]

\footnotetext{For our example, the instrumentation, coverage measurement and
  coverage report can be reproduced with the artifact at \artifacturl}



\bluecov{} emits a \emph{coverage report} in JSON format which shows
the information of Table~\ref{tab:coverage-report-ftest-example}. The
results for our example are as expected: the condition in line 3 is
executed for each unit test and therefore the associated two goals are
hit 3 times each. The condition in line 5 and the associated two goals
are reached 2 times and the last test in line 8 is reached one time
only. Each return statement in line 4, 6 and 8 is reached exactly
once. And finally the return statement in line 9 is not reached at
all.

\small

\begin{table}[th]
  \centering
  \begin{tabular}[h]{lc}
    \toprule
    goal name & hit count \\
    \midrule
    \verb|FloatTools.sign:(F)I.coverage.1| & 3 \\
    \verb|FloatTools.sign:(F)I.coverage.2| & 3 \\
    \verb|FloatTools.sign:(F)I.coverage.3| & 2 \\
    \verb|FloatTools.sign:(F)I.coverage.4| & 1 \\
    \verb|FloatTools.sign:(F)I.coverage.5| & 2 \\
    \verb|FloatTools.sign:(F)I.coverage.6| & 1 \\
    \verb|FloatTools.sign:(F)I.coverage.7| & 1 \\
    \verb|FloatTools.sign:(F)I.coverage.8| & 1 \\
    \verb|FloatTools.sign:(F)I.coverage.9| & 0 \\
    \bottomrule
  \end{tabular}
  \caption{Coverage Report for \emph{FloatTools.sign}}
  \label{tab:coverage-report-ftest-example}
\end{table}

\normalsize

\subsection{Enhance Test Coverage}
\label{sec:enhance-coverage}

At a first glance, it seems that line 9 is simply dead code which cannot be
covered, but, when telling \jbmc{} to cover the coverage goal 9, \jbmc{} does in fact
report that it can complete the test-suite because it found a way to execute
line 9.%

The input value for this is an IEEE-754 ``not a number'' (\emph{NaN})
value. This is a valid value for the input parameter \emph{x} and \emph{NaN}
values have the property that each comparison with such a value evaluates to
\emph{false}. Therefore, we can complete the test suite by adding the
unit test shown in listing~\ref{lst:additional-test-case}.

\begin{lstlisting}[caption=Addtitionally Generated Unit Test\label{lst:additional-test-case}]
  @Test
  public void testSignNaN () {
    assertEquals(-2, FloatTools.sign(Float.NaN));
  }
\end{lstlisting}

After adding this additional test and re-executing the full
test-suite, \bluecov{} reports coverage of all location coverage goals of
\emph{FloatTools.sign}.

\section{Bytecode Instrumentation and Coverage Measurement}
\label{sec:bytec-instr}

To perform the bytecode instrumentation in \bluecov{} we chose the
ASM~\footnote{\url{https://asm.ow2.io/}} library. ASM is widely used to
instrument Java bytecode, e.g., in
JaCoCo~\footnote{\url{https://www.eclemma.org/jacoco/}}. It uses the visitor
pattern for instrumentation.

The challenge here is being able to instrument the bytecode at the
right positions and to reliably get the information at runtime which
instrumented bytecode is executed. In particular, it is necessary to
make sure that even non-standard program termination like uncaught
exceptions or direct calls to exit via JNI do not prevent the
information from being stored persistently so that it can be reported
back to \jbmc{} for test generation.

\subsection{Bytecode Instrumentation}
\label{sec:impl-instr}

Each bytecode instruction is visited and can be changed. It is interesting to
note that ASM does not provide the information about the byte offset or address
of an instruction. As instructions exist which can have different sizes, e.g.,
`iload 0' and its specialization `iload\_0'. Instead, we identify the bytecode
instructions by their bytecode index.  The bytecode index is the sequence number
of bytecode instructions and is independent of the size in bytes. \jbmc{} provides
the bytecode index for each coverage goal.

Each bytecode instruction specified for instrumentation by the coverage
criterion of \jbmc{} is instrumented by inserting the following code directly
before the bytecode instruction where \emph{uid} is the unique identifier of the
instrumented location.

\begin{lstlisting}
companyCoverageReporter.record(uid);
\end{lstlisting}

This Java code is translated into bytecode by pushing the static field
\emph{companyCoverageReporter} on the stack, then the uid of the instrumented
location and finally calling the \emph{record} method of the
\emph{org.cprover.coverage.CoverageLog} class.

This instrumentation is performed as shown in
listing~\ref{lst:bytecode-instrumentation}. Here we show the visiting method for
jump instructions. Analogous methods exist for each of the other categories of
bytecode instructions as well. It calls directly \emph{instrumentByteCode}
method with the current bytecode index as argument which checks whether the
given bytecode index should be instrumented. If yes, the necessary bytecode
instructions and arguments are inserted into the code, if no, nothing is
done. On return to \emph{visitJumpInsn}, the bytecode index counter is
incremented and the next bytecode instruction is visited. In this way, all
bytecode instructions are visited and each one that corresponds to a coverage
goal gets instrumented with the necessary code.

\begin{lstlisting}[caption=Bytecode Instrumentation Implementation\label{lst:bytecode-instrumentation}]
public final void visitJumpInsn(final int opcode,
                                final Label label) {
  instrumentByteCode(bcLine);
  bcLine += 1;
  super.visitJumpInsn(opcode, label);
}

final void instrumentByteCode(final int bcLine) {
  if (shouldBeInstrumented(bcLine)) {
    lastMethodWasInstrumented = true;
    // get instance from static field
    // push value to record
    // call `record` on CoverageLog
    super.visitFieldInsn(Opcodes.GETSTATIC,
        this.className,
        "companyCoverageReporter",
        "Lorg/cprover/coverage/CoverageLog;");
    super.visitLdcInsn(getUniqueIdentifier(bcLine));
    super.visitMethodInsn(Opcodes.INVOKEVIRTUAL,
        "org/cprover/coverage/CoverageLog",
        "record",
        "(I)V",
        false);
    debug("added ID " + getUniqueIdentifier(bcLine));
    instrumentedLocs.add(getUniqueIdentifier(bcLine));
  }
}
\end{lstlisting}

\begin{figure*}[ht]
  \centering
  \includegraphics[width=0.6\textwidth]{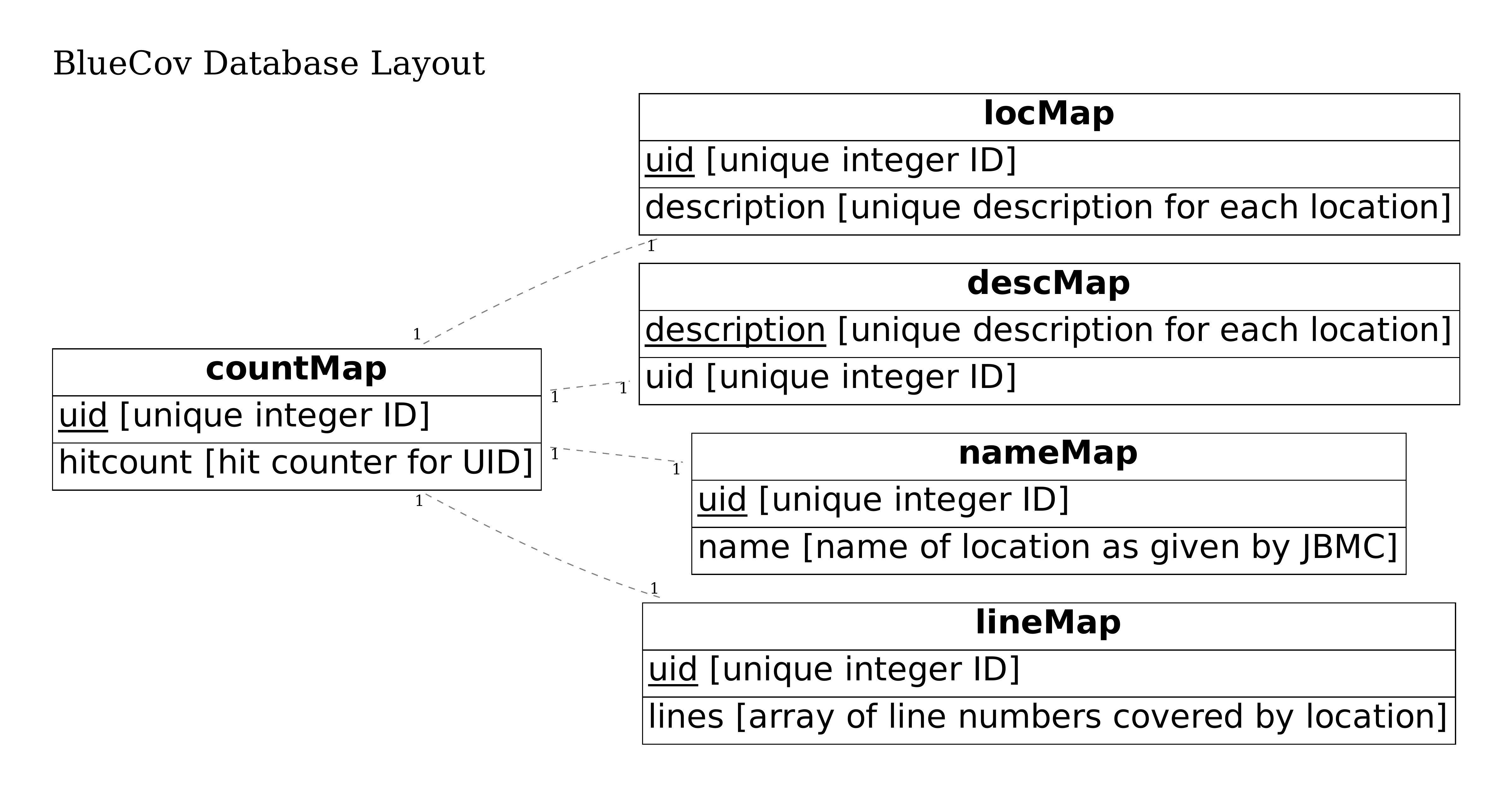}
  \caption{MapDB HTree objects}
  \label{fig:mapdb-layout}
\end{figure*}

\subsection{\bluecov{} Runtime Coverage Measurement}
\label{sec:runtime-coverage}

At runtime, whenever an instrumented bytecode instruction is executed,
there is a call to \bluecov{} to register the coverage of the
associated location. This increments the hit count of the
corresponding location. This happens in the \emph{CoverageLog} class
of the \emph{org.cprover.coverage} package. The \emph{record} method
is shown in listing~\ref{lst:record-coverage-goal}.

\begin{lstlisting}[caption=Recording Reaching a Coverage Goal\label{lst:record-coverage-goal}]
public void record(final int key) {
  Integer i = inMemoryMap.get(key);
  if (i == null) {
    inMemoryMap.put(key, 1);
  } else {
    inMemoryMap.put(key, i + 1);
  }
}
\end{lstlisting}

Storing this information persistently is realized using the \emph{mapDB}
library\footnote{\url{https://mapdb.org/}}. This library provides a combination
of Java collections and on-disk database storage, i.e., there is a thread-safe
hash map which is in-memory and also mirrored on disk. The ER diagram of the
relevant part of the database is shown in~\ref{fig:mapdb-layout}.


\subsubsection{Challenges}
\label{sec:challenges}

Using exact coverage measurement provides some challenges which had to be
addressed. For each bytecode instruction which is instrumented for the coverage
analysis, there exists one call to get the \emph{CoverageLog} singleton object
and one call to record execution of the corresponding bytecode instruction. The
required time to complete this has to be minimized in order to keep execution
time of the project under analysis reasonable.

To achieve this, \bluecov{} uses an in-memory database in its normal execution
mode and only writes the in-memory content to the \emph{mapDB} on-disk database
on exit. This greatly reduces the time required to execute the projects under
test.

The drawback here is that it has to be guaranteed that the in-memory database is
correctly written at each possible program termination. This is achieved by
installing a \emph{shutdown hook} for the JVM runtime using the
\emph{addShutdownHook} method of the \emph{Runtime} class. This method takes a
\emph{Thread} as parameter, its \emph{run} method get called on normal exit but
also if the JVM terminates due to a user action or system-wide
event\footnote{\url{https://docs.oracle.com/javase/8/docs/api/java/lang/Runtime.html\#addShutdownHook(java.lang.Thread)}}. When
the hook is called the code in listing~\ref{lst:inmem-db-to-disk} is executed
and writes the in-memory content to the on-disk database on exit.

\begin{lstlisting}[caption=Write In-Memory DB to Disk\label{lst:inmem-db-to-disk}]
  public void run() {
    if (inMemory) {
      db = makeDb();
      countMap = db.hashMap(locCountMap)
          .keySerializer(Serializer.INTEGER)
          .valueSerializer(Serializer.INTEGER)
          .createOrOpen();
      for (Integer key : inMemoryMap.keySet()) {
        Integer orig = countMap.get(key);
        if (orig == null) {
          orig = 0;
        }
        countMap.put(
            key, orig + inMemoryMap.get(key));
      }
    }
    db.close();
  }
\end{lstlisting}

\subsection{Java Project Integration}
\label{sec:java-proj-integr}

\bluecov{} can easily be integrated into Maven projects. This requires adding
the necessary runtime libraries to the dependencies section of the
\emph{pom.xml} file. The \emph{bluecov.jar} file location is specified in the
plugins section of the \emph{pom.xml} file. In this way, the coverage analysis
is used when Maven runs the test suite.

\section{Conclusion}
\label{sec:conclusion}

We presented an approach to focussing a Java test generator based on bounded
model checking on the uncovered test goals of an existing test suite. \bluecov{}
achieves this by performing a tailored bytecode instrumentation that captures
exactly the coverage goals that correspond to a chosen coverage metric provided
by the test generation tool.

This avoids mismatches in the coverage criteria implemented in existing coverage
measurement tools and thus avoids the generation of redundant tests. This direct
integration of formal analysis and test coverage analysis allows for reduction
of cost of automated test case generation and an increase in precision. The
minimization of newly generated tests increases clarity of the test suite which
is important for human testers interacting with the test suite.

As open-source software, \bluecov{} is freely available and can be integrated
into Java projects. Together with \jbmc{} this can help in analyzing and
completing test suite coverage for projects.


\paragraph{Related work} The CProver framework has been used to implement several test generation
tools based on bounded model checking, mainly for C programs, the most
versatile one being FShell~\cite{DBLP:conf/vmcai/HolzerSTV09}, which
offers a domain specific language for expressing arbitrary coverage
goals.
The tool ChainCover~\cite{DBLP:conf/pts/SchrammelMK13,DBLP:journals/sttt/SchrammelMK16} aims at reducing the initialization overhead of testing
reactive systems by merging tests into test scenarios.

Semi-automated approaches to closing test coverage
gaps~\cite{DBLP:conf/nfm/NellisKCKST16} have been considered as well
as automated ones, e.g.~\cite{DBLP:conf/qsic/BloemKRT14}, but none if
these works considers the problem of filling coverage gaps by
exactly integrating with runtime coverage measurement in order to avoid the
generation of tests that overlap with existing automatically or
manually created test suites.

\paragraph{Future work} The proposed approach can be also taken to focus model
checking on untested code, similarly to conditional model checking~\cite{DBLP:conf/sigsoft/BeyerHKW12} except that the conditions are not based on
information from another verification engine, but a coverage analysis tool.



\hide{
\footnotesize
\begin{minted}[linenos]{C}
FloatTools.sign(float) /* FloatTools.sign:(F)I */
  DECL return_tmp0 : floatbv[32]
  // FloatTools.java line 5 FloatTools.sign:(F)I bytecode-index 1
  CALL java.lang.Math.<clinit>()
  // FloatTools.java line 5 FloatTools.sign:(F)I bytecode-index 1
  CALL java.lang.Math.abs:(F)F(x)
  // FloatTools.java line 5 FloatTools.sign:(F)I bytecode-index 1
  ASSIGN return_tmp0 := java.lang.Math.abs:(F)F::return_value
  // FloatTools.java line 5 FloatTools.sign:(F)I bytecode-index 1
  DEAD java.lang.Math.abs:(F)F::return_value
  // FloatTools.java line 5 FloatTools.sign:(F)I bytecode-index 5
  IF (NOT isnan(return_tmp0)) AND (return_tmp0 < 1.000000e-6)) THEN GOTO 1
  // FloatTools.java line 7 FloatTools.sign:(F)I bytecode-index 8
  DEAD return_tmp0
  // FloatTools.java line 7 FloatTools.sign:(F)I bytecode-index 8
  GOTO 2
  1: DEAD return_tmp0
  // FloatTools.java line 6 FloatTools.sign:(F)I bytecode-index 7
  ASSIGN FloatTools.sign:(F)I::return_value := 0
  // FloatTools.java line 6 FloatTools.sign:(F)I bytecode-index 7
  GOTO 5
  // FloatTools.java line 7 FloatTools.sign:(F)I bytecode-index 11
  2: IF isnan(x) OR (x >= 0) THEN GOTO 3
  // FloatTools.java line 8 FloatTools.sign:(F)I bytecode-index 13
  ASSIGN FloatTools.sign:(F)I::return_value := -1
  // FloatTools.java line 8 FloatTools.sign:(F)I bytecode-index 13
  GOTO 5
  // FloatTools.java line 9 FloatTools.sign:(F)I bytecode-index 17
  3: IF isnan(x) OR (x <= 0) THEN GOTO 4
  // FloatTools.java line 10 FloatTools.sign:(F)I bytecode-index 19
  ASSIGN FloatTools.sign:(F)I::return_value := 1
  // FloatTools.java line 10 FloatTools.sign:(F)I bytecode-index 19
  GOTO 5
  // FloatTools.java line 11 FloatTools.sign:(F)I bytecode-index 21
  4: ASSIGN FloatTools.sign:(F)I::return_value := -2
  5: END_FUNCTION
\end{minted}
\normalsize
}

\hide{
\subsection{Testcase Minimization}
\label{sec:testc-minim}

In many applications \jbmc{} will have to create multiple new test cases in order
to complete the test suite. Often there are different possibilities to cover the
remaining coverage goals. As each generated unit test might cover several
coverage goals, it makes sense to try to minimize the number of generated unit
tests to complete the coverage.

The underlying problem is the subset cover problem which is
$\mathbb{NP}$-complete and therefore is not solved in an optimal way. \jbmc{} uses
a heuristic to approximate the minimal number of new unit tests.

For algorithm~\ref{alg:test-case-minimization}, let $S$ be the set of traces for
the coverage as computed by \jbmc{}.\@ This set is converted into a list which is
sorted in descending order by the number of goals covered by each trace. The
output of the algorithm is a set of traces $T\subseteq S$. At each iteration,
the next trace $t$ is examined: if it covers a previously uncovered goal, then
$t$ is added to $T$, else it is dropped.

\begin{algorithm}[ht]
  \caption{Test Case Minimization Heuristic}
  \label{alg:test-case-minimization}
  \begin{algorithmic}
    \State $l \gets to\_list(S);$
    \State $l \gets sort\_ascending(l);$
    \State $T \gets \emptyset;$
    \State $G \gets \emptyset;$
    \While{$l\neq Nil$}
    \State $t \leftarrow head(l);$ \Comment{first element of list l}
    \State $l \leftarrow tail(l);$ \Comment{rest of list l}
    \If{$\neg(goals(t) \subseteq G)$} \Comment{does the trace cover a new goal not yet in G?}
    \State $G \gets G \cup goals (t);$
    \State $T \gets T \cup \{t\};$
    \EndIf
    \EndWhile
  \end{algorithmic}
\end{algorithm}
}

\end{document}